\documentclass[preprint2]{aastex61}

\newcommand\aastex{AAS\TeX}

\usepackage{epstopdf}

\received{March 19, 2017}
\revised{March 18, 2017}
\accepted{\today}

\submitjournal{ApJ}

\shorttitle{\aastex\ Large-amplitude longitudinal oscillations in a solar filament}
\shortauthors{Zhang et al.}

\begin{document}

\title{Large-amplitude longitudinal oscillations in a solar filament}

\correspondingauthor{Qingmin Zhang}
\email{zhangqm@pmo.ac.cn}

\author[0000-0003-4078-2265]{Q. M. Zhang}
\affil{Key Laboratory for Dark Matter and Space Science, Purple Mountain Observatory, CAS, Nanjing 210008, China}
\affil{CAS Key Laboratory of Solar Activity, National Astronomical Observatories, Beijing 100012, China}

\author{T. Li}
\affil{CAS Key Laboratory of Solar Activity, National Astronomical Observatories, Beijing 100012, China}
\affil{University of Chinese Academy of Sciences, Beijing 100049, China}

\author{R. S. Zheng}
\affil{Institute of Space Sciences, Shandong University, Weihai 264209, China}

\author{Y. N. Su}
\affil{Key Laboratory for Dark Matter and Space Science, Purple Mountain Observatory, CAS, Nanjing 210008, China}

\author{H. S. Ji}
\affil{Key Laboratory for Dark Matter and Space Science, Purple Mountain Observatory, CAS, Nanjing 210008, China}

\begin{abstract}
In this paper, we report our multiwavelength observations of the large-amplitude longitudinal oscillations of a filament on 2015 May 3. Located next to active region 12335, the sigmoidal filament was observed by 
the ground-based H$\alpha$ telescopes from the Global Oscillation Network Group (GONG) and by the Atmospheric Imaging Assembly (AIA) instrument aboard the \textit{Solar Dynamics Observatory} (\textit{SDO}). 
The filament oscillations were most probably triggered by the magnetic reconnection in the filament channel, which is characterized by the bidirectional flows, brightenings in EUV and SXR, and magnetic cancellation 
in the photosphere. The directions of oscillations have angles of 4$^\circ$-36$^\circ$ with respect to the filament axis. The whole filament did not oscillate in phase as a rigid body. Meanwhile, the periods (3100$-$4400 s) of 
oscillations have a spatial dependence, implying that the curvature radii ($R$) of the magnetic dips are different at different positions. The values of $R$ are estimated to be 69.4$-$133.9 Mm, and the minimum transverse 
magnetic field of the dips is estimated to be 15 G. The amplitudes of S5-S8 grew with time, while the amplitudes of S9-S14 damped with time. The amplitudes of oscillations range from a few to ten Mm, and the 
maximal velocity can reach 30 km s$^{-1}$. Interestingly, the filament experienced mass drainage southwards at a speed of $\sim$27 km s$^{-1}$. The oscillations continued after the mass drainage and lasted for more than 11 hr. 
After the mass drainage, the phases of oscillations did not change a lot. The periods of S5-S8 decreased, while the periods of S9-S14 increased. The amplitudes of S5$-$S8 damped with time, while the amplitudes 
of S9-S14 grew. Most of the damping (growing) ratios are between -9 and 14. We propose a schematic cartoon to explain the complex behaviors of oscillations by introducing thread-thread interaction.
\end{abstract}


\section{Introduction} \label{sec:intro}

Solar prominences are cool and dense plasmas in the hot corona \citep{tan95,mar98,lab10,mac10,par14}. Long-term H$\alpha$ observations of the full disk from the ground-based
telescopes have considerably improved our understandings of prominence. On the solar disk, they are dark features with parallel fine threads, which are called filament.
Filaments are located along the polarity inversion lines (PILs) either in the active regions or in the quiet region, even near the polar region \citep{su12,su15}.
It is generally believed that prominences/filaments are suspended by the upward magnetic tension force of the sheared arcades or magnetic flux ropes \citep{aul98,dev00,yan15}. 
The formation of cool material in the filament channels is partly associated with chromospheric evaporation and the subsequent condensation along the dip as a result of thermal instability \citep{kar06,luna12b,zhou14}.
Despite of being stable during most of their lifetimes, they are full of dynamics, such as counter-streaming \citep{zir98,sch10,chen14}, magneto-thermal convection \citep{ber08,ber11}, and oscillations \citep{lin11}.
When the filaments are disturbed or the condition of a certain kind of instability is fulfilled \citep{tor04,kli06,li15}, they will probably become unstable and erupt, generating flares \citep{ji03,zqm15,zqm17}, 
jets \citep{ster15,zqm16}, and coronal mass ejections (CMEs) \citep{zheng17}.

Filament oscillations are ubiquitous in the solar corona \citep[][and references therein]{oli02,arr12}. According to the velocity amplitude, they can be classified into small-amplitude oscillations ($\le$3 km s$^{-1}$) 
and large-amplitude oscillations ($\ge$20 km s$^{-1}$). The former type is often caused by the propagation of magnetohydrodymic (MHD) waves with periods of a few minutes \citep[e.g.,][]{lin07,oka07,ning09}. 
The later type, however, is often caused by incoming disturbances, such as Moreton waves and EUV waves at speeds of $\sim$1000 km s$^{-1}$ \citep[e.g.,][]{eto02,gil08,asai12}. 
The direction of large-amplitude filament oscillation could be horizontal \citep{kle69} or vertical to \citep{hyd66,ram66,kim14} the photosphere. The amplitudes of displacement of the filament axis range from several to 
tens of Mm \citep{dai12,gos12}. The period ranges from a few minutes to hours \citep{iso06,shen14a}. The amplitudes of oscillations always damp with time, and the damping timescales are tens of minutes \citep{bocc11,her11}. 
The detected periods are useful for indirect diagnostics of magnetic field strength \citep{pou06} and the angles between the magnetic field and the filament main axis \citep{reg01}.

Apart from the large-amplitude transverse oscillations, there are large-amplitude longitudinal oscillations in filaments. For the first time, \citet{jing03} observed filament oscillation along the axis using the H$\alpha$ 
observations from the Big Bear Solar Observatory (BBSO). Triggered by a subflare near a footpoint of the filament, the oscillation lasted for $\ge$4 hr. The velocity amplitude, period, and damping time are 92 km s$^{-1}$, 
80 minutes, and 210 minutes, 
respectively. \citet{vrs07} observed similar periodic motion along the filament. Based on a magnetic flux rope model, the authors proposed that the oscillation is triggered by the poloidal magnetic flux injection by magnetic 
reconnection at one of the filament legs and the restoring force is magnetic pressure gradient along the filament axis. Using the high-resolution and high-cadence Ca~{\sc ii} H (3968.5 {\AA}) observations of the active region
(AR) 10940 from the Solar Optical Telescope \citep[SOT;][]{tsu08} aboard the \textit{Hinode} spacecraft, \citet{zqm12} studied the longitudinal prominence oscillation, which lasted for $\ge$3.5 hr on 2007 February 8. 
The initial velocity, period, and damping timescale are 40 km s$^{-1}$, 52 minutes, and 133 minutes, respectively.
Combing the observation and one-dimensional (1D) hydrodynamic (HD) numerical simulation of prominence oscillations, they found that the restoring force is the component of gravity along the dip, which is supported by 
the following analytical solutions and numerical simulations \citep[e.g.,][]{luna12a,luna12c,zqm13,luna16a,luna16b}. A simple expression of the period of longitudinal oscillation is derived, i.e., $P=2\pi \sqrt{R/g_{\odot}}$, 
where $R$ and $g_{\odot}$ represent the curvature radius of the magnetic dip and the surface gravity acceleration of the Sun. Therefore, it is instructive to estimate the curvature radius and the lower limit of 
the transverse magnetic field strength of the dip according to the observed period \citep[e.g.,][]{bi14,luna14,pant16}.

Unlike in the transverse oscillations, mass drainage occasionally takes place in the large-amplitude longitudinal oscillations. 
For the first time, \citet{bi14} reported that a significant amount of filament material was drained toward the southern filament endpoint at a speed of 62 km s$^{-1}$ after the oscillation. The authors concluded that the 
mass drainage plays an important role in the transition from the slow rise to fast rise of the eruptive filament. Actually, the mass drainage has been predicted by the numerical simulation when the initial velocity exceeds 
a critical value \citep[see Fig. 8 in][]{zqm13}. The remaining material continues to oscillate with the same period, although the damping time decreases remarkably. The continuing longitudinal oscillations after mass drainage, 
however, have never been observed and reported. What are the differences of filament oscillations before and after mass drainage?
Moreover, the growing amplitudes of longitudinal filament oscillations have not been noticed before.  In this paper, we report our multiwavelength observations 
of the large-amplitude longitudinal oscillation of a filament on 2015 May 3, focusing on its initiation, mass drainage, and the continuing oscillation. Data analysis is described in Section~\ref{sec:data}. Results are presented 
in Section~\ref{sec:result}. Many aspects of longitudinal filament oscillations are discussed in Section~\ref{sec:discuss}. Finally, we give a summary of the results in Section~\ref{sec:summary}.

\section{Instruments and data analysis} \label{sec:data}

Located near NOAA AR 12335 (S14E47), the intermediate filament was continuously observed by the Global Oscillation Network Group (GONG) in H$\alpha$ line center (6562.8 {\AA})
and the Atmospheric Imaging Assembly \citep[AIA;][]{lem12} 
aboard the \textit{Solar Dynamics Observatory} (\textit{SDO}) in EUV wavelengths (171, 304, and 335 {\AA}). The photospheric line-of-sight (LOS) magnetograms were observed by the Helioseismic and Magnetic 
Imager \citep[HMI;][]{sch12} aboard \textit{SDO}. The level\_1 data from AIA and HMI were calibrated using the standard \textit{Solar Software} (\textit{SSW}) programs \textit{aia\_prep.pro} and \textit{hmi\_prep.pro}. 
The full-disk H$\alpha$ and AIA 304 {\AA} images are coaligned with an accuracy of $\sim$1$\farcs$2 using the cross correlation method. The large-scale three-dimensional (3D) magnetic configuration near the filament 
was derived from the potential field source surface \citep[PFSS;][]{sch03} modeling. The soft X-ray (SXR) flux in 1$-$8 {\AA} on that day was recorded by the \textit{GOES} spacecraft.
The observational parameters, including the instrument, wavelength, time, cadence, and pixel size are summarized in Table~\ref{tab:para}.

\begin{deluxetable}{ccccc}
\tablecaption{Description of the observational parameters \label{tab:para}}
\tablecolumns{5}
\tablenum{1}
\tablewidth{0pt}
\tablehead{
\colhead{Instru.} &
\colhead{$\lambda$} &
\colhead{Time} & 
\colhead{Cad.} & 
\colhead{Pix. size} \\
\colhead{} & 
\colhead{({\AA})} &
\colhead{(UT)} & 
\colhead{(s)} & 
\colhead{(\arcsec)}
}
\startdata
GONG & 6562.8 & 12:07$-$23:59 & 60 & 1.0 \\
\textit{SDO}/AIA & 171 & 12:00$-$23:59 & 24\tablenotemark{a} & 0.6 \\
\textit{SDO}/AIA & 304 & 12:00$-$23:59 & 24\tablenotemark{a} & 0.6 \\
\textit{SDO}/AIA & 335 & 12:00$-$23:59 & 24\tablenotemark{a} & 0.6 \\
\textit{SDO}/HMI & 6173 & 11:00$-$14:00 & 45 & 0.5 \\
\textit{GOES} & 1$-$8 & 11:00$-$14:00 & 2.05 & \nodata \\
\enddata
\tablenotetext{a}{To reduce workload, we use AIA data with a cadence of 24 s, although the original cadence is 12 s.}
\end{deluxetable}

\section{Results} \label{sec:result}

\subsection{Magnetic field and configuration} \label{sec:mag}
\begin{figure*}
\plotone{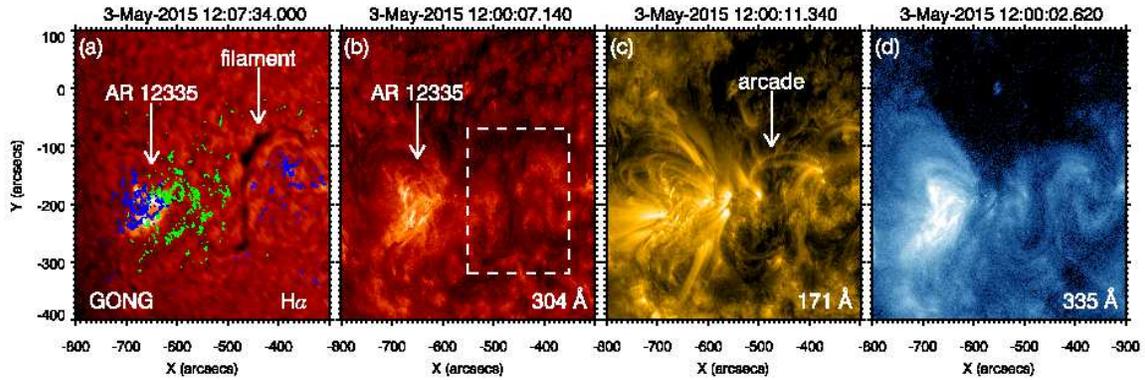}
\caption{The AR and filament observed in H$\alpha$, 304, 171, and 335 {\AA}. 
In panel (a), the contours ($\pm$200, $\pm$300, and $\pm$400 G) of positive and negative LOS magnetic field at 12:07:21 UT are superposed with green and blue lines, respectively. 
In panel (b), the white dashed box is used to calculate the integral EUV intensities around the filament. In panel (c), the arrow points at the hot arcade overlying the filament.
\label{fig1}}
\end{figure*}

Figure~\ref{fig1} shows the bright AR and dark filament observed in H$\alpha$, 304, 171, and 335 {\AA} before oscillation. In panel (a), the AR with the highest intensity is pointed by the white arrow.
The contours of positive and negative LOS magnetic field at 12:07:21 UT are superposed with green and blue lines, respectively. It is obvious that the sigmoidal filament is located along the PIL. In panel (c), the 171 {\AA} 
image indicates that the filament is constrained by the overlying right-skewed arcade with temperature of $\sim$1 MK.

\begin{figure*}
\plotone{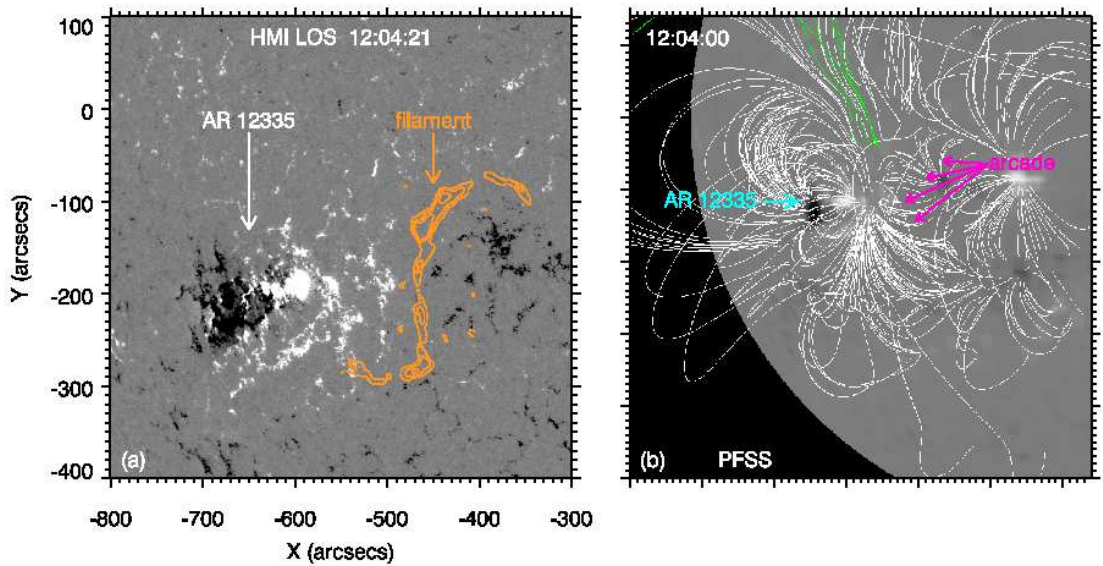}
\caption{\textit{Left:} HMI LOS magnetogram at 12:04:21 UT.
The intensity contours of the filament at 12:07:34 UT are superposed with orange lines.
\textit{Right:} 3D magnetic configuration near the filament at 12:04:00 UT using the PFSS modeling. Closed and open magnetic field lines are drawn with white and green lines. 
The cyan and magenta arrows point at the AR and arcade, respectively.
\label{fig2}}
\end{figure*}

In Figure~\ref{fig2}, the left panel demonstrates the HMI LOS magnetogram at 12:04:21 UT. The intensity contours of the filament at 12:07:34 UT are superposed with orange lines, showing that the long filament ($\sim$370$\arcsec$ 
in length) is located between the positive and negative polarities. 
The 3D magnetic configuration around the filament using the PFSS modeling is displayed in the right panel. It is clear that the magnetic arcade overlying the filament is consistent 
with that in the AIA 171 {\AA} image (see Figure~\ref{fig1}(c)). 

\subsection{Triggering of the filament oscillations} \label{sec:trig}
\begin{figure*}
\epsscale{.80}
\plotone{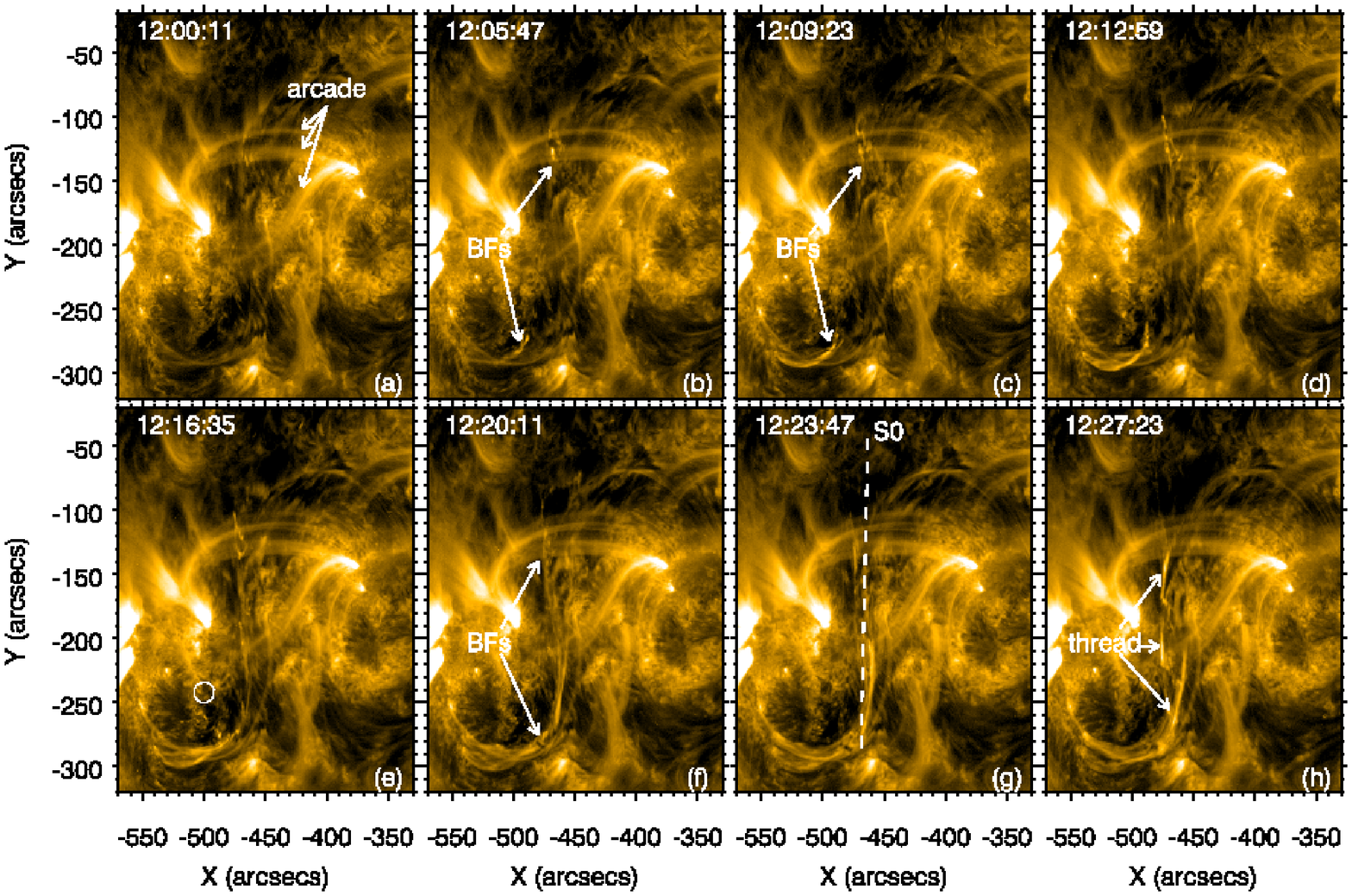}
\caption{Eight snapshots of the EUV 171 {\AA} images during 12:00$-$12:27 UT. In panels (b), (c), and (f), ``BFs'' means bidirectional flows.
In panel (e), the white circle indicates the position of magnetic cancellation.
In panel (g), the dashed line (S0) is used for investigating the magnetic reconnection before the filament oscillations.
In panel (h), the white arrows point at the bright, fine threads in the filament channel.
(An animation of this figure is available.)
\label{fig3}}
\end{figure*}

\begin{figure}
\epsscale{0.85}
\plotone{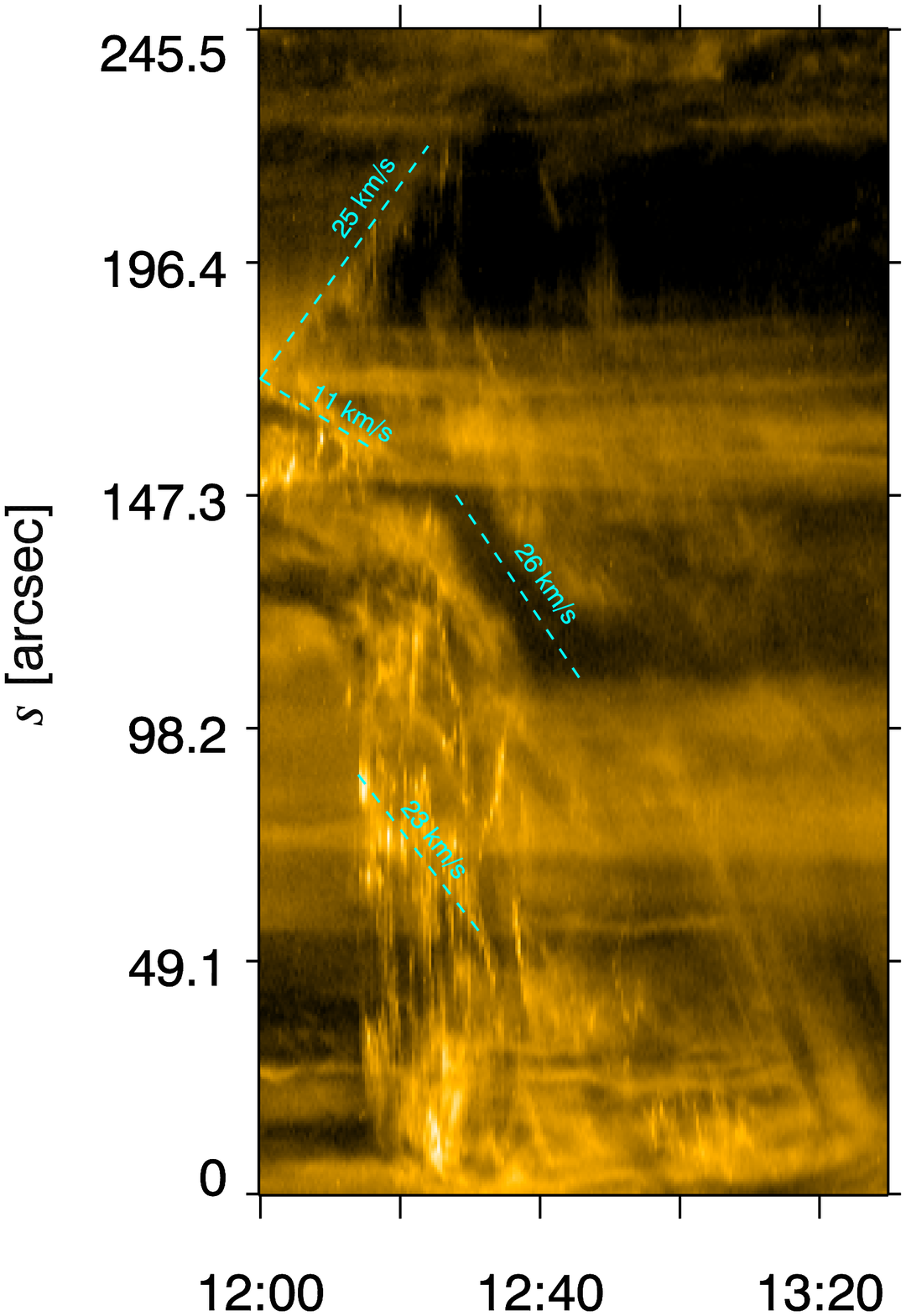}
\caption{Time-slice diagram of S0 in 171 {\AA}. $s=0\arcsec$ and $s=245\farcs5$ in the $y$-axis represent the southern and northern endpoints, respectively.
The speeds of bidirectional flows and filament are labeled next to the cyan dashed lines.
\label{fig4}}
\end{figure}

Before the filament oscillation, there was plausible evidence of magnetic reconnection. Figure~\ref{fig3} shows eight snapshots of the EUV images in 171 {\AA} during 12:00$-$12:27 UT (see also the online animated figure).
The filament was quiet and stable before 12:00 UT.
After $\sim$12:05:00 UT, bidirectional flows suddenly appeared in the filament channel, which are pointed by the white arrows in panel (b). The plasmas moved northwards and southwards simultaneously, lasting for 
$\ge$30 minutes. In order to investigate the bidirectional flows, we extract the intensities along a straight line S0, which is plotted in panel (g) 
with a long dashed line. The temporal evolution of the intensities along S0 in 171 {\AA} is illustrated with the time-slice diagram in Figure~\ref{fig4}, where $s=0\arcsec$ and $s=245\farcs5$ represent the southern and northern 
endpoints, respectively. The bidirectional flows are obviously shown, with the slopes of the inclined features stand for the speeds of outflows \citep[e.g.,][]{yang15}. The speed of northward flow is $\sim$25 km s$^{-1}$, 
while the speeds of southward flows are $\sim$11 and $\sim$23 km s$^{-1}$. The dark filament started to move southwards at $\sim$12:28 UT at a speed of $\sim$26 km s$^{-1}$.

\begin{figure*}
\plotone{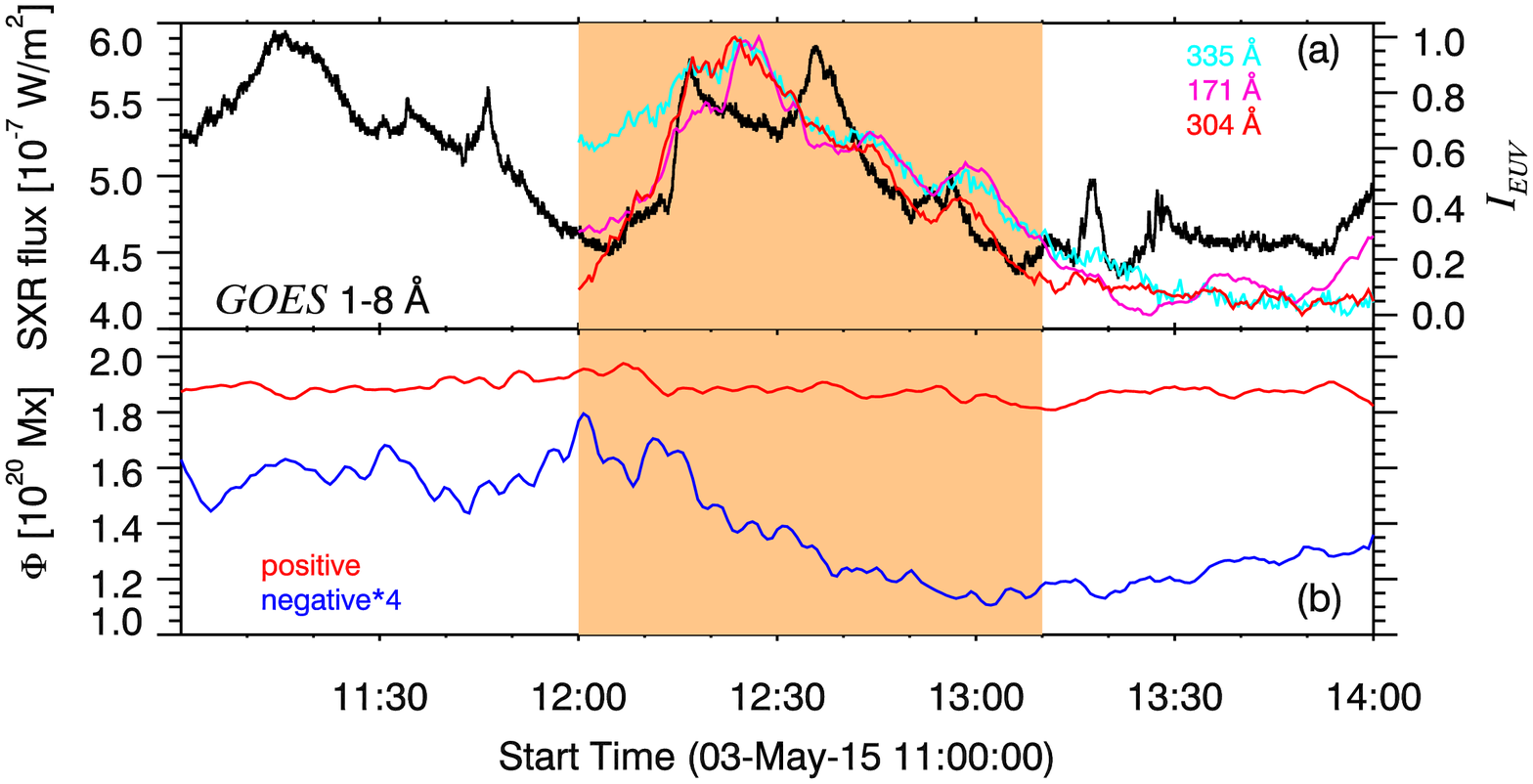}
\caption{(a) \textit{GOES} SXR light curve in 1$-$8 {\AA} (\textit{black}) and normalized EUV light curves around the filament channel 
in 171 {\AA} (\textit{magenta}), 304 {\AA} (\textit{red}), and 335 {\AA} (\textit{cyan}).
(b) Temporal evolutions of the unsigned positive (\textit{red}) and negative (\textit{blue}) magnetic fluxes within the field-of-view (FOV) of Figure~\ref{fig6}.
\label{fig5}}
\end{figure*}

In typical solar flares and microflares, free magnetic energy is released and converted into the thermal and kinetic energies of plasmas via magnetic reconnection, and the intensities in SXR 
and EUV increase significantly. For this event, we observed brightenings of the threads (see Figure~\ref{fig3}). We sum up the EUV intensities within the dashed box of Figure~\ref{fig1}(b). 
In Figure~\ref{fig5}, the temporal evolutions of the normalized EUV intensities in 171, 304, and 335 {\AA} are plotted with magenta, red, and cyan lines, respectively. The EUV intensities increased rapidly 
from $\sim$12:05 UT and reached the maxima at $\sim$12:25 UT before decreasing gradually until $\sim$13:30 UT. We also plot the \textit{GOES} SXR light curve in 1$-$8 {\AA} with a black solid line. It is obvious 
that the EUV intensities have close correlations with the SXR light curve, implying that the increased SXR emission originates mainly from the bright threads along the filament channels during that time. We also 
calculated the integral EUV intensities of AR 12335, which are absolutely inconsistent with the SXR light curve. It should be emphasized that the brightening of the threads is so weak that it only amounts to a B6.0 flare, 
i.e., a microflare.

\begin{figure*}
\epsscale{1.0}
\plotone{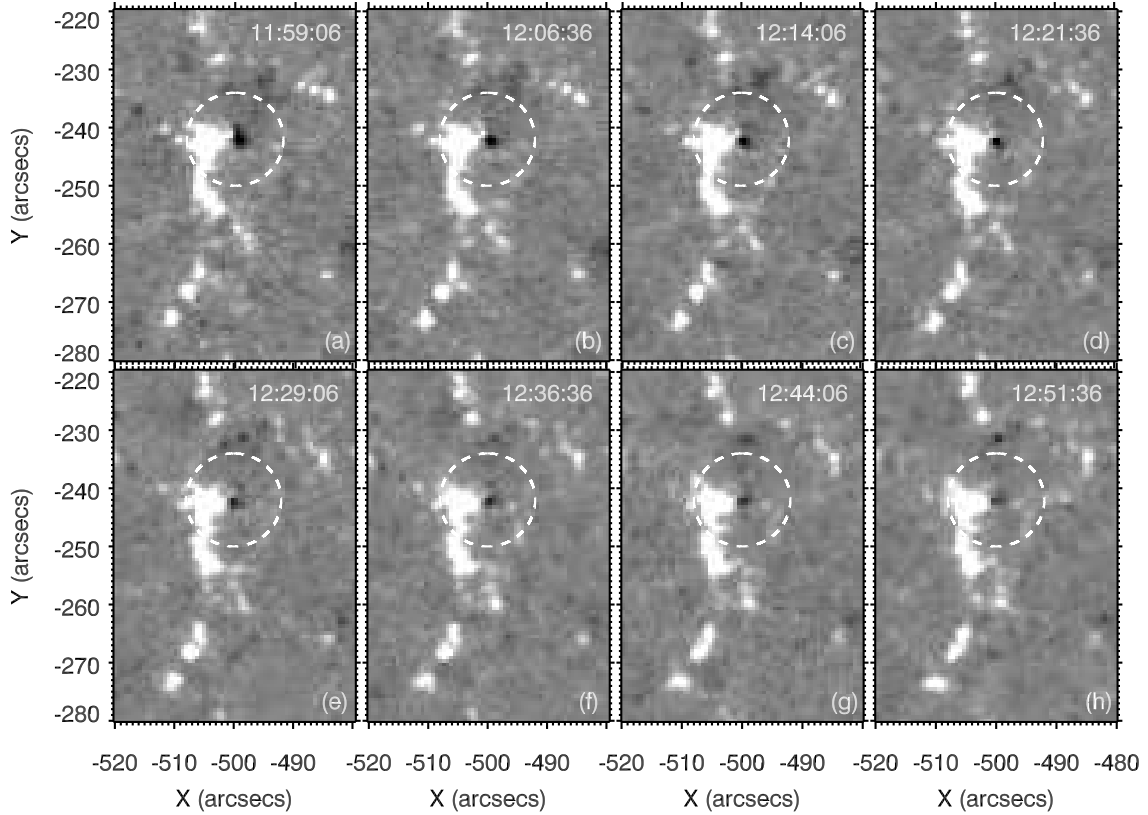}
\caption{Eight snapshots of the HMI LOS magnetograms during 11:59$-$12:52 UT. The dashed circles indicate where the magnetic cancellation occurs.
\label{fig6}}
\end{figure*}

Magnetic reconnection in the chromosphere or corona are always associated with variation of the photospheric magnetic field, such as emerging flux and magnetic cancellation \citep[e.g.,][]{ster05,su11,chen15,liu16}. 
Figure~\ref{fig6} shows eight snapshots of the LOS magnetograms during 11:59$-$12:52 UT. After careful inspection, we found that a tiny negative polarity was cancelled,
which is indicated by the white dashed circles. During 12:00$-$13:00 UT, the area and strength of the negative polarity adjacent to the filament channel decreased gradually. We calculated the unsigned positive 
and negative fluxes within the whole FOV. The temporal evolutions of the fluxes are plotted with red and blue lines in Figure~\ref{fig5}(b). It is revealed that the negative flux decreases gradually during 12:00$-$13:00 UT,
which is coincident with the period of brightenings of the threads (see Figure~\ref{fig5}(a)). Actually, the positive flux also decreases by $\sim$1.5$\times$10$^{19}$ Mx at the same time, which is approximately the 
same amount as that of the negative flux. In Figure~\ref{fig3}(e), we label the region of magnetic cancellation with a solid circle, which is very close to the source region of the bidirectional flows. 
Therefore, we conclude that the magnetic cancellation is temporally and spatially associated with the brightenings of the threads heated by magnetic reconnection.

\subsection{Filament oscillations} \label{sec:osci}
\begin{figure*}
\epsscale{1.0}
\plotone{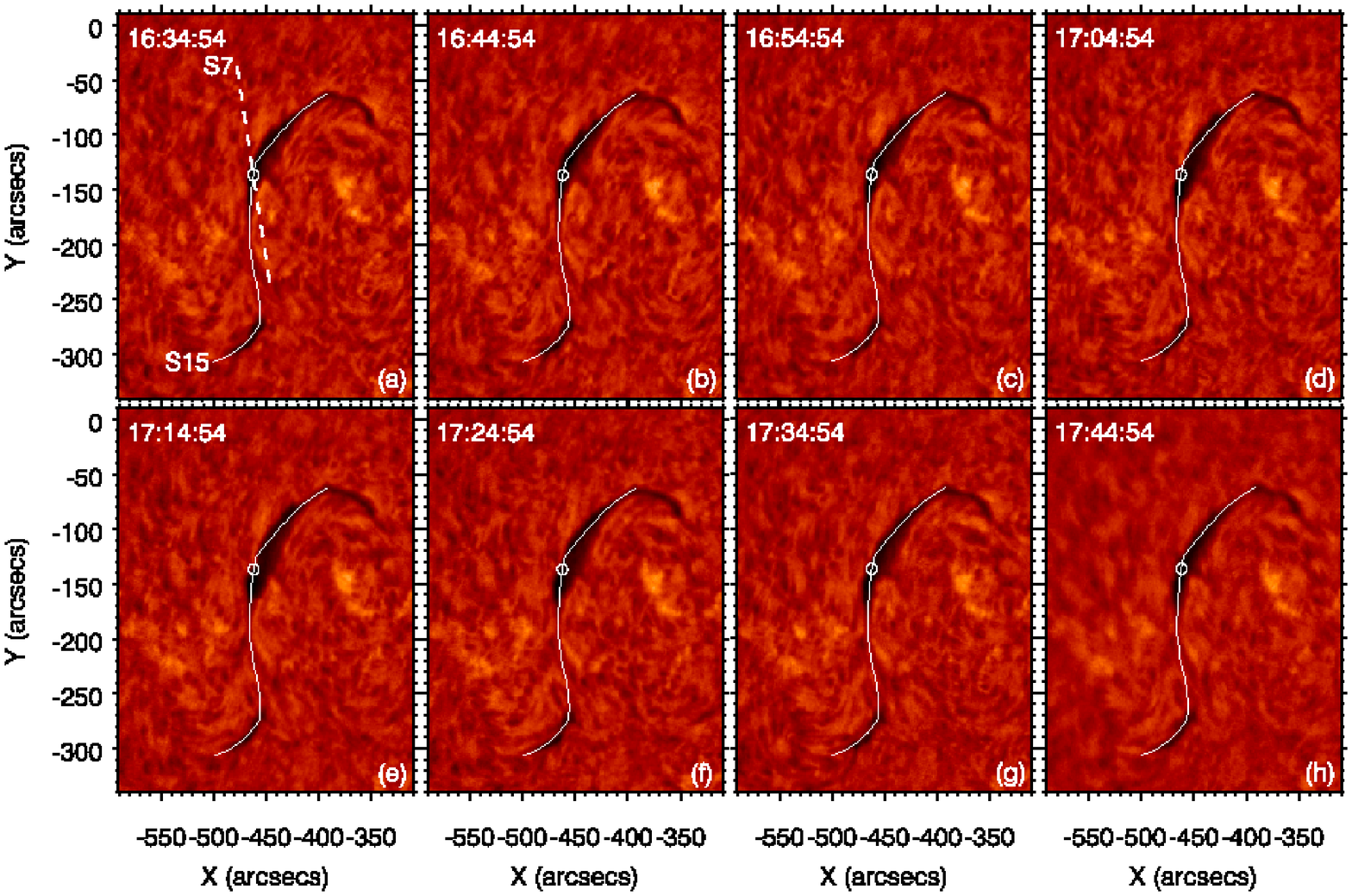}
\caption{Eight snapshots of the H$\alpha$ images during 16:34$-$17:44 UT.
(An animation of this figure is available.)
\label{fig7}}
\end{figure*}

After the magnetic reconnection and brightening in the filament channel, the filament oscillated for more than 11 hr.
The eight snapshots of the H$\alpha$ images in Figure~\ref{fig7} show an example of one cycle of filament oscillation, where the spine (S15) derived before the oscillation is fixed (see also the online animated figure).
It is obvious that the dark filament material moved southwestward from $\sim$16:34 UT until $\sim$17:15 UT. Then, the material moved backwards to its initial position until $\sim$17:45 UT.

\begin{figure*}
\epsscale{1.0}
\plotone{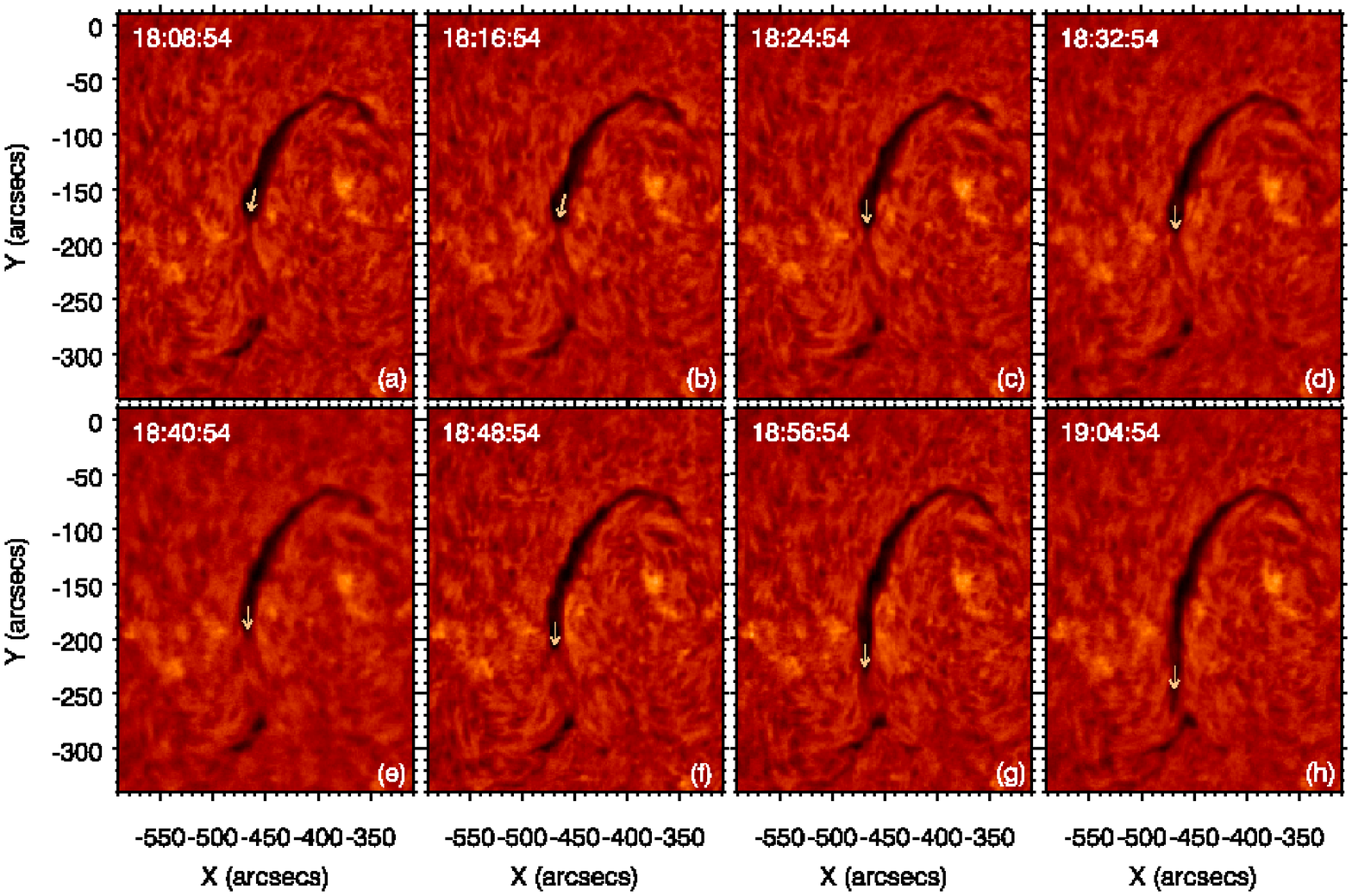}
\caption{Eight snapshots of the H$\alpha$ images during 18:08$-$19:05 UT.
The yellow arrows indicate the directions of mass drainage along the filament spine.
(An animation of this figure is available.)
\label{fig8}}
\end{figure*}

After a few cycles, the filament material underwent mass drainage. Figure~\ref{fig8} shows eight snapshots of the H$\alpha$ images where the yellow arrows indicate the directions of mass drainage.
It is clear that the material moved southwards along the filament channel during 18:08$-$19:05 UT.

\begin{figure}
\epsscale{1.0}
\plotone{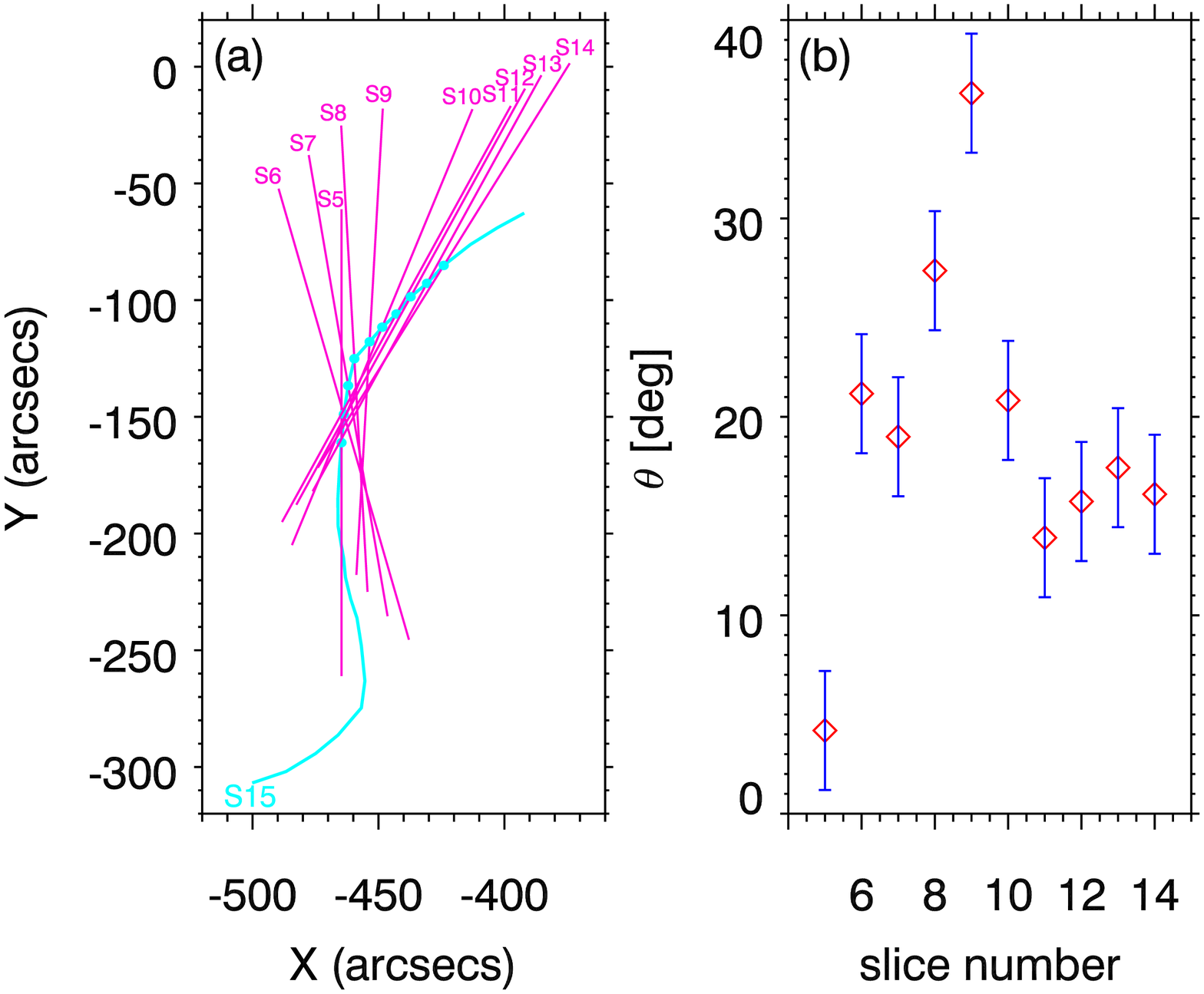}
\caption{(a) The spine (S15) of the filament (\textit{cyan}) and ten slices (S5$-$S14) passing through the filament (\textit{magenta}).
(b) The angles ($\theta$ in degree) between the ten slices and spine.
\label{fig9}}
\end{figure}

From the online animations, we found that only part of the filament oscillated, instead of the whole body. In Figure~\ref{fig9}(a), We selected ten positions along the oscillating segment. Using the same 
method as that in \citet{luna14}, for each position, we extracted the H$\alpha$ and EUV intensities of 31 straight lines along 31 directions ranging from 30$^\circ$ to 120$^\circ$ with respect to the EW direction.
We made time-slice diagrams along the straight lines and compared the oscillation patterns. The direction with the maximum amplitude is considered to be the real direction of oscillation. 
In Figure~\ref{fig9}(a), the ten magenta lines (S5$-$S14 with length of 200\arcsec) represent the directions of oscillation at the ten positions. The angles ($\theta$) between the magenta lines and the spine are displayed 
in Figure~\ref{fig9}(b). It is revealed that most of the angles lie in the range of 14$^\circ$-28$^\circ$, with exceptionally low value of 4.2$^\circ$ and high value of 36.3$^\circ$. Therefore, the filament oscillations are 
longitudinal. In the event of longitudinal filament oscillations on 2010 August 20, most of the angles between the spine and directions of oscillations lie in the range of 15$^\circ$-40$^\circ$ \citep{luna14}, which are 
close to our results.

\begin{figure*}
\epsscale{.80}
\plotone{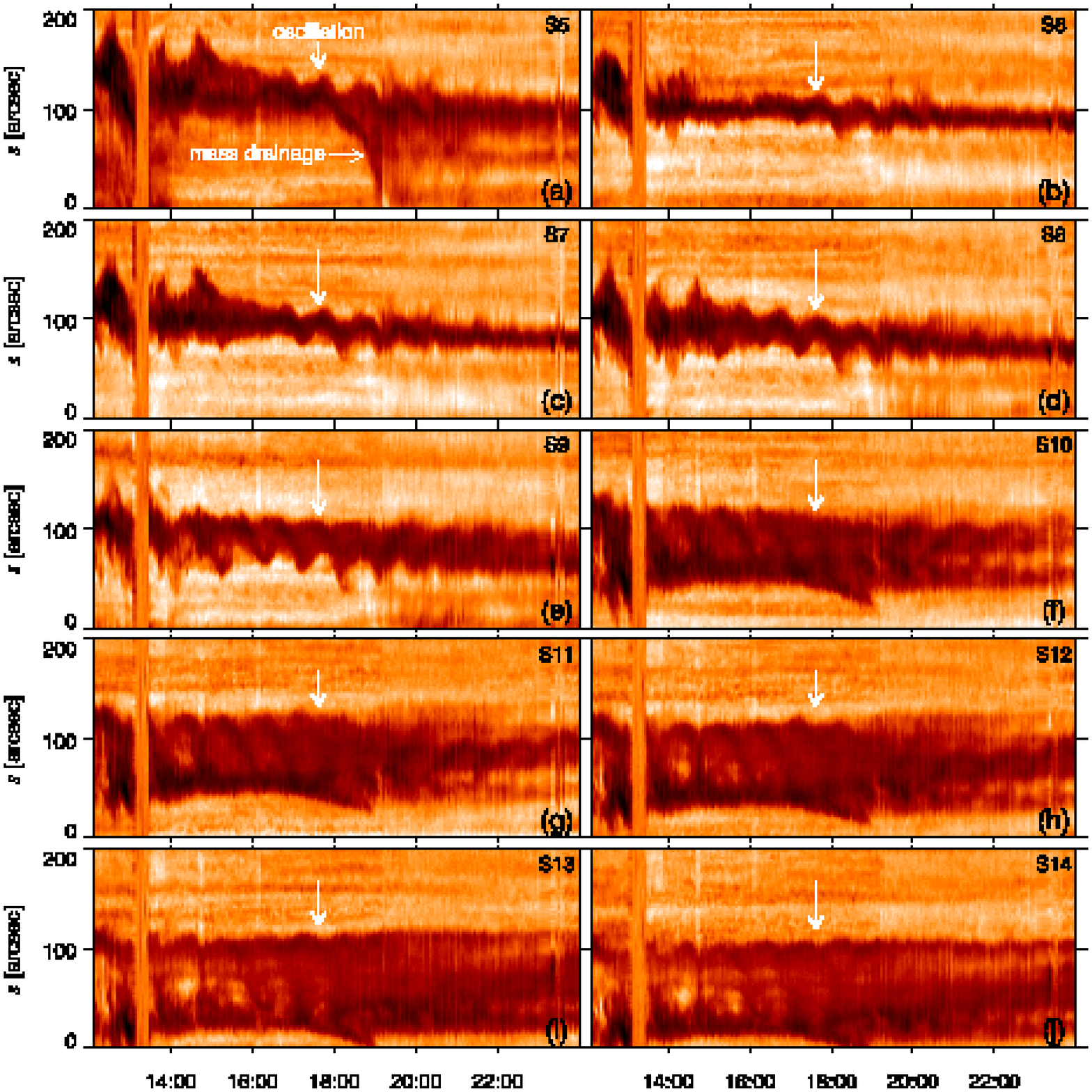}
\caption{Time-slice diagrams of S5$-$S14 in H$\alpha$. $s=0\arcsec$ and $s=200\arcsec$ in the $y$-axis represent the southern and northern endpoints of the slices. 
In panel (a), the white arrow points at the southwards mass drainage. The thick white arrows point at the filament oscillations.
\label{fig10}}
\end{figure*}

The time-slice diagrams of S5$-$S14 in H$\alpha$ are drawn in Figure~\ref{fig10}. It is evident that oscillations are present in all the diagrams. In panel (a), there is a signature of mass drainage during 18:08$-$19:05 UT, which 
is consistent with the H$\alpha$ images in Figure~\ref{fig8}. Interestingly, the oscillation did not cease after the drainage. Instead, the remaining part continued to oscillate, which has never been reported in previous literatures.
Owing to the lower resolution and cadence of the H$\alpha$ images, the oscillation patterns for S10$-$S14 are less clearer than S5$-$S9. Due to the possible superposition of multi-components along the line of sight, 
the apparent trajectories of S10$-$S14 in Figure~\ref{fig10}(f-j) look larger than those of S5$-$S9 in Figure~\ref{fig10}(a-e).

\begin{figure*}
\epsscale{.80}
\plotone{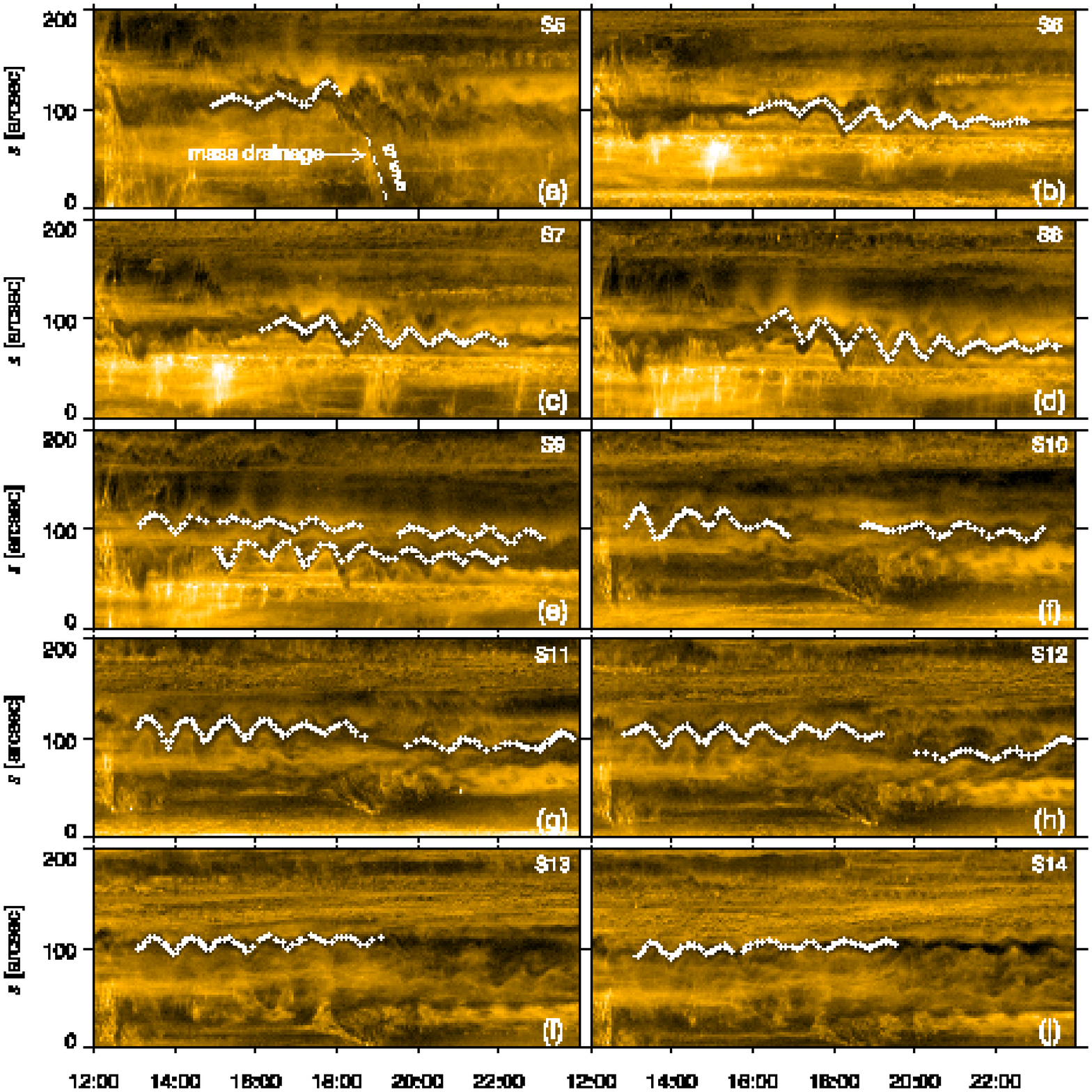}
\caption{Time-slice diagrams of S5$-$S14 in 171 {\AA}. $s=0\arcsec$ and $s=200\arcsec$ in the $y$-axis represent the southern and northern endpoints of the slices.
In panel (a), the white arrow points at the southwards mass drainage at the speed of $\sim$27 km s$^{-1}$. We mark the positions of filament manually with white ``+'', 
which are used for curve fitting in Figure~\ref{fig12}.
\label{fig11}}
\end{figure*}

Similar to Figure~\ref{fig10}, Figure~\ref{fig11} shows the time-slice diagrams of the ten slices in 171 {\AA}. Thanks to the higher resolution and cadence of AIA, the filament oscillations are much more striking than in H$\alpha$.
The oscillations present very complex behaviors. On one hand, the amplitudes of oscillations of S5$-$S8 grew with time before the mass drainage. Afterwards, the amplitudes damped with time. 
On the other hand, the amplitudes of S9$-$S14 damped with time initially and increased after mass drainage. It is noticed that there are upper and lower wave trains with almost antiphases along S9, 
implying multiple oscillating filament threads (see panel (e)). Besides, the starting times of oscillations for S5$-$S8 lag behind those of S9$-$S14 by approximately two hours.

\begin{figure*}
\epsscale{.80}
\plotone{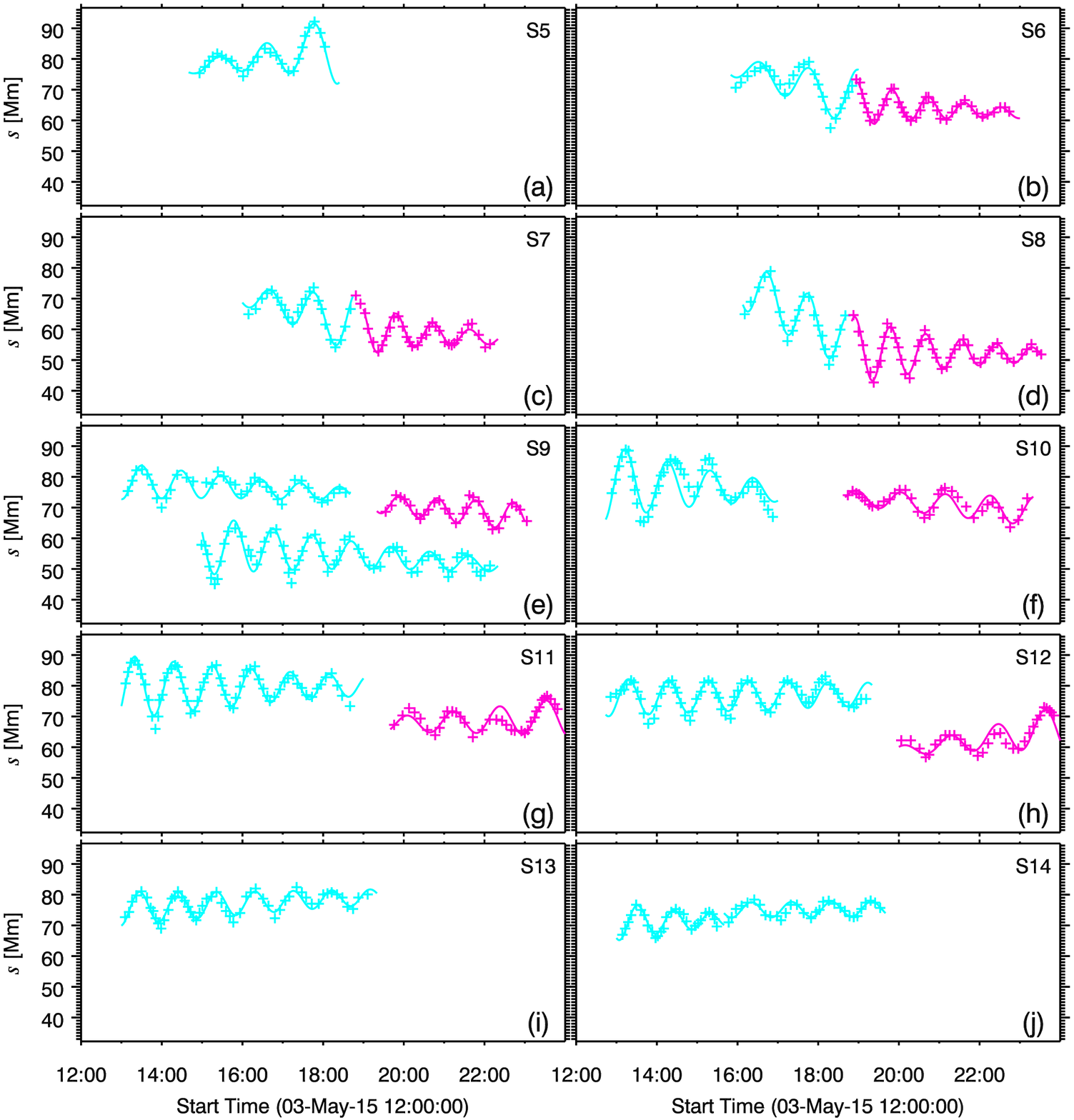}
\caption{Fitted curves of the longitudinal filament oscillations before (\textit{cyan}) and after (\textit{magenta}) the mass drainage.
\label{fig12}}
\end{figure*}

In order to calculate the parameters of the longitudinal filament oscillations, we mark the positions of filament manually with white ``+'' in Figure~\ref{fig11} and fit the curves using the widely adopted function 
\citep[e.g.,][]{jing03,vrs07,zqm12}:
\begin{equation} \label{eqn1}
y=y_0+bt+A\sin(\frac{2\pi}{P}t+\phi)e^{-t/\tau},
\end{equation}
where $y_0$, $A$, and $\phi$ signify the initial position, amplitude, and phase at 12:00 UT.
$b$, $P$, and $\tau$ stand for the linear velocity of the material, period, and damping (growing) timescale of oscillation, respectively. The amplitudes of oscillations are damping (growing) if $\tau$ is positive (negative).
In Figure~\ref{fig12}, we plot the results of curve fitting using the standard program \textit{mpfit.pro} in \textit{SSW}. The cyan and magenta lines represent the results before and after the mass drainage, respectively. 
It is obvious that the function in Equation~\ref{eqn1} can perfectly describe the filament oscillations. The amplitudes of oscillations range from a few to ten Mm, and the amplitudes of S5$-$S9 are slightly larger 
than those of S10$-$S14. The fitted parameters, including $\phi$, $P$, and $\tau$, are listed in the middle three columns of Table~\ref{tab:fit}. The calculated damping (growing) ratios ($\tau/P$) are listed in the last column.

\begin{figure*}
\epsscale{.80}
\plotone{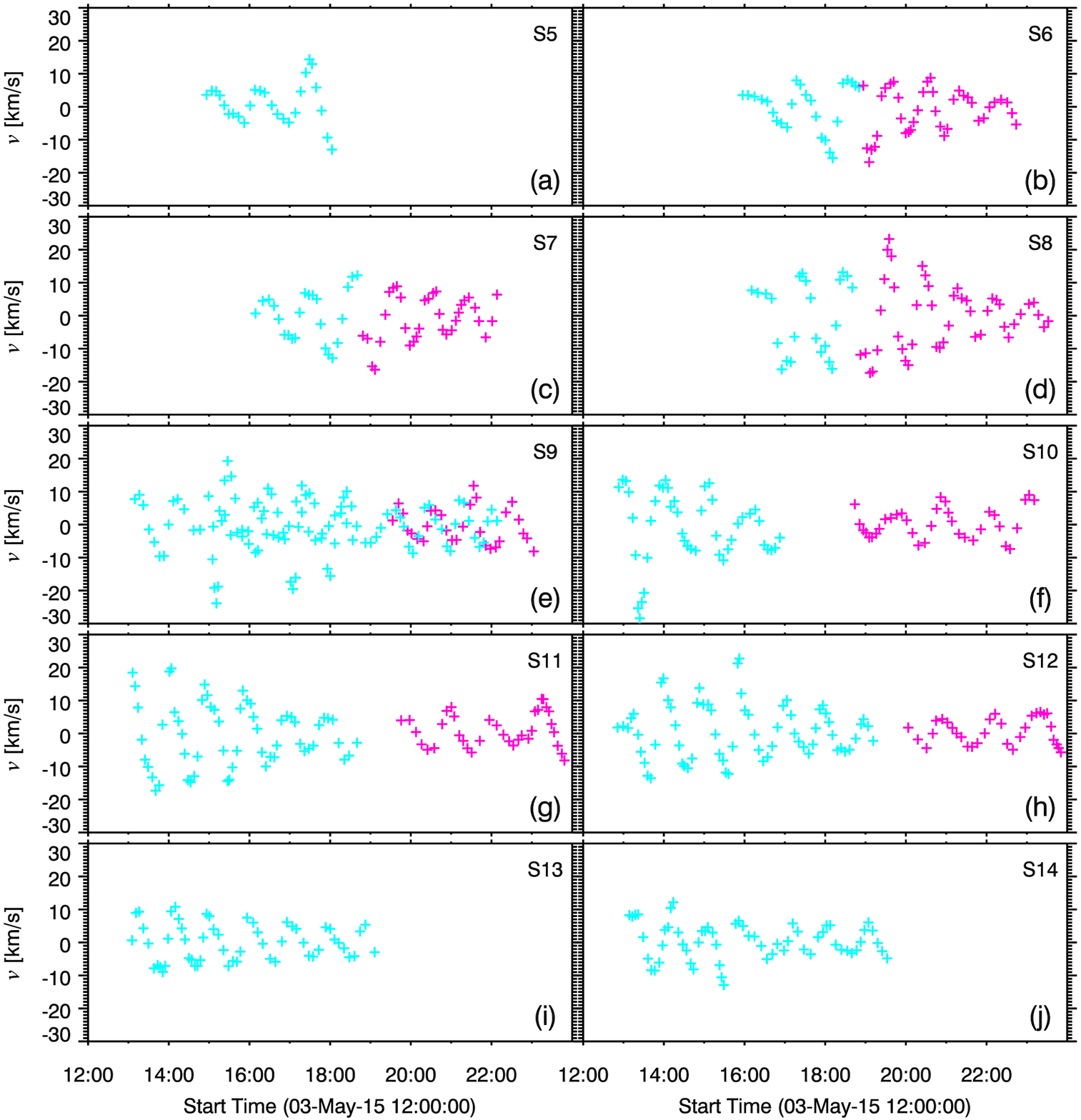}
\caption{Velocities of the longitudinal filament oscillations before (\textit{cyan}) and after (\textit{magenta}) the mass drainage.
\label{fig13}}
\end{figure*}

Based on the displacements of the filament in Figure~\ref{fig12}, we calculated the velocities of the filament oscillations ($v=ds/dt$). The results of calculation are displayed in Figure~\ref{fig13}. It is clear that the 
velocities of the filament range from 0 to 30 km s$^{-1}$, although most of the values are less than 20 km s$^{-1}$.

\begin{deluxetable}{ccccc}
\tablecaption{Fitted parameters of the longitudinal filament oscillations \label{tab:fit}}
\tablecolumns{5}
\tablenum{2}
\tablewidth{0pt}
\tablehead{
\colhead{Slice} & 
\colhead{$\phi$} &
\colhead{$P$} & 
\colhead{$\tau$} &
\colhead{$\tau/P$} \\
\colhead{} &  
\colhead{(rad)} &
\colhead{(second)} & 
\colhead{(second)} & 
\colhead{}
}
\startdata
S5 &  -0.96 & 4221.50 & -6643.58 & -1.57 \\
S6 &  0.22 & 4392.44 & -9488.63 & -2.16 \\
S6 &  0.17 & 3234.76 & 8968.83 &  2.77 \\
S7 &  2.69 & 3892.72 & -8192.06 & -2.10 \\
S7 &  1.42 & 3299.90 & 7555.92 & 2.29 \\
S8 &  0.47 & 3647.18 & -88302.90 & -24.21 \\
S8 &  -0.73 & 3162.51 & 10477.20 & 3.31 \\
S9 &  1.37 & 3534.62 & 19656.30 & 5.56 \\
S9 &  2.24 & 3386.17 & -30292.90 & -8.95 \\
S9 &  -0.97 & 3495.27 & 16461.10 & 4.71 \\
S10 & 3.77 & 3809.15 & 11387.30 & 2.99 \\
S10 & 2.94 & 3978.81 & -25054.50 & -6.30 \\
S11 & 2.30 & 3468.91 & 18338.90 & 5.29 \\
S11 & 4.54 & 4135.59 & -16907.70 & -4.09 \\
S12 & 2.75 & 3559.89 & 48121.20 & 13.52 \\
S12 & -1.09 & 4211.97 & -9133.60 & -2.17 \\
S13 & 1.42 & 3429.04 & 27612.80 & 8.05 \\
S14 & 0.35 & 3245.81 & 8304.86 & 2.56 \\
S14 & -0.81 & 3573.13 & 373353.00 & 104.49 \\
\enddata
\end{deluxetable}

\begin{figure}
\epsscale{1.0}
\plotone{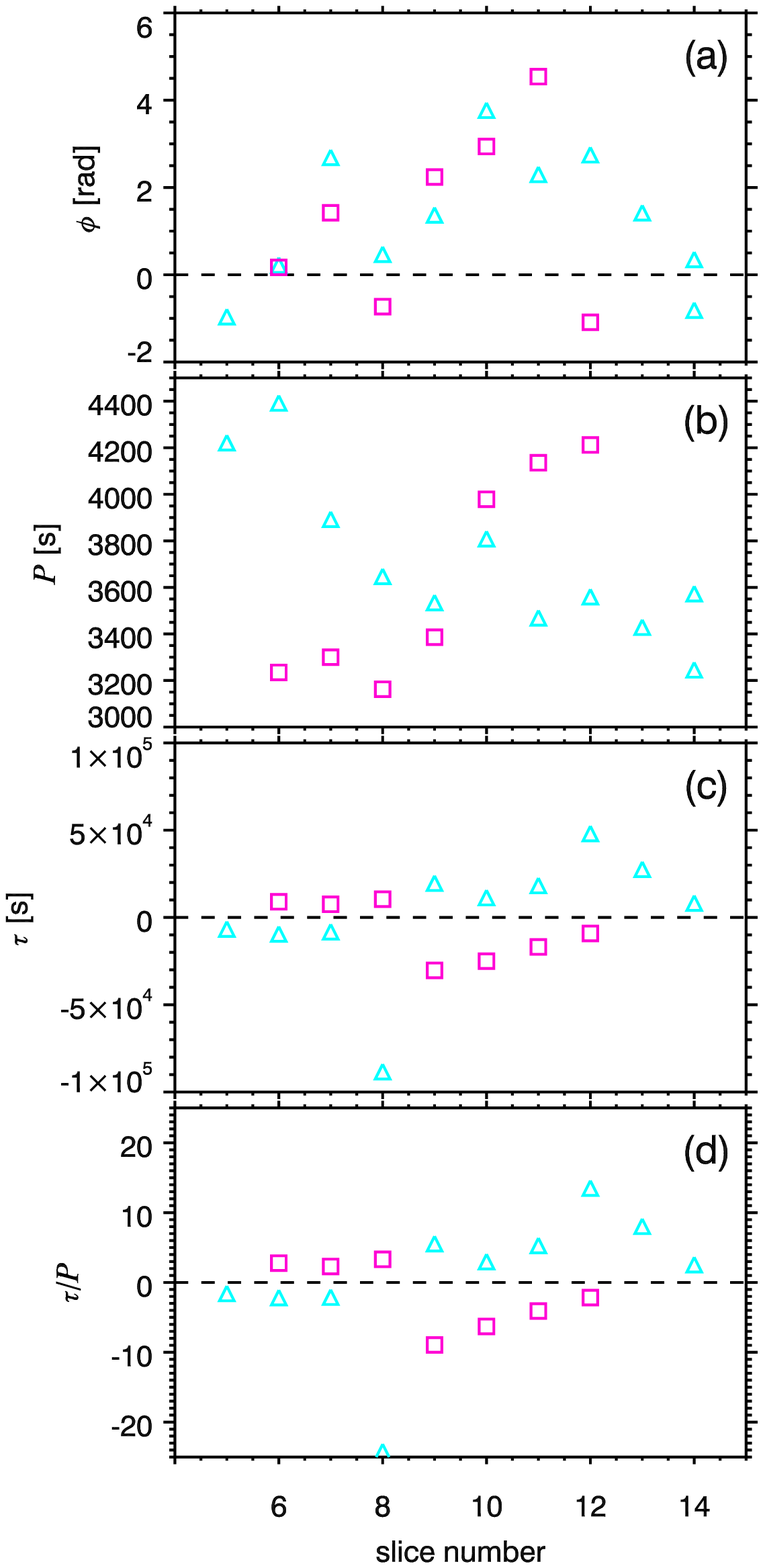}
\caption{(a-c) Parameters of the filament oscillations, including the initial phase, period, damping (growing) timescale.
The damping (growing) ratios are calculated and plotted in panel (d).
The cyan and magenta symbols represent the results before and after the mass drainage.
\label{fig14}}
\end{figure}

In order to analyze the results more conveniently, we plot the parameters as a function of slice number in Figure~\ref{fig14}. In panel (a), it is clear that the initial phases are different, ranging from -1 to 4.5, which means 
that the filament did not oscillate in phase as a solid body. On the other hand, the phases did not change a lot after the mass drainage for most of the positions (slices) except S11 and S12, which is in agreement 
with the prediction of numerical simulation \citep{zqm13}. There are two oscillations at two apparently different locations along S11 and S12. Therefore, the phases and other parameters do not neccessarily match 
before and after the mass drainage. In panel (b), the periods of oscillation range from 3100 s to 4400 s, which are longer than the values reported by \citet{vrs07,zqm12,luna14}, but lower than the values reported 
by \citet[][]{jing03,shen14b,wang16}. Interestingly, the periods decreased after mass drainage for S6-S9, but increased after mass drainage for S10-S12, which has never been reported before. 
In panel (c), most of the damping (growing) timescales lie in the range of -3$\times$10$^4$ s to 5$\times$10$^4$ s, which are in the same order of magnitude as the previous values \citep{jing03,jing06,vrs07,zqm12,luna14}. 
It is interesting that the amplitudes of oscillations increased before the mass drainage and decreased afterwards for S6-S8. On the other hand, the amplitudes decreased before the mass drainage and 
increased afterwards for S9-S12. The extraordinarily large values of $\tau$ (-88302.90 s and 373353.00 s) in Table~\ref{tab:fit} suggest that the amplitude almost keeps constant without damping or growing,
which has been noticed in previous studies. On 2012 April 7, after longitudinal oscillation without damping, the southern part of a filament erupted and generated a flare/CME event \citep{li12}.
In panel (d), most the damping (growing) ratios are in the range of -9 to 14.

\section{Discussion} \label{sec:discuss}
\subsection{How are the filament oscillations triggered?}
Large-amplitude filament oscillations have been observed and investigated for many decades. The transverse oscillations can be triggered by EUV waves \citep[e.g.,][]{asai12,dai12,gos12}, 
Moreton waves \citep[e.g.,][]{eto02,gil08,liu13}, shock waves \citep{shen14b,pant16}, and magnetic reconnection between the emerging flux and the magnetic arcade overlying the filament \citep{chen08}.
For the longitudinal oscillations, the triggering mechanisms are mostly subflares (microflares) near the footpoints of filaments \citep{jing03,vrs07,zqm13}, episodic jets \citep{luna14}, and shock wave \citep{shen14b,pant16}.
In our study, the filament oscillations are triggered by a $\sim$B6.0 microflare within the filament channel. As described in Section~\ref{sec:trig}, the magnetic reconnection is characterized by the bidirectional flows 
(see Figure~\ref{fig3}), EUV and SXR brightenings (see Figure~\ref{fig5}(a)), and magnetic cancellation in the photosphere (see Figure~\ref{fig5}(b) and Figure~\ref{fig6}). The magnetic reconnection plays a key role in the 
reconfiguration of the magnetic topology, which serves as a disturbance to the filament. Therefore, longitudinal filament oscillations are triggered in the directions that have angles of 4$^\circ$-36$^\circ$ with respect to the axis 
(see Figure~\ref{fig9}). As illustrated in the cartoon of \citet{shen14b}, the direction of filament oscillation depends closely on the direction of incoming disturbance. In this study, the bidirectional flows within the filament channel 
are consistent with the longitudinal oscillations. Hence, the oscillations could not be transverse.

\subsection{Curvature radii and magnetic field strength of the dips}
\begin{figure}
\epsscale{1.0}
\plotone{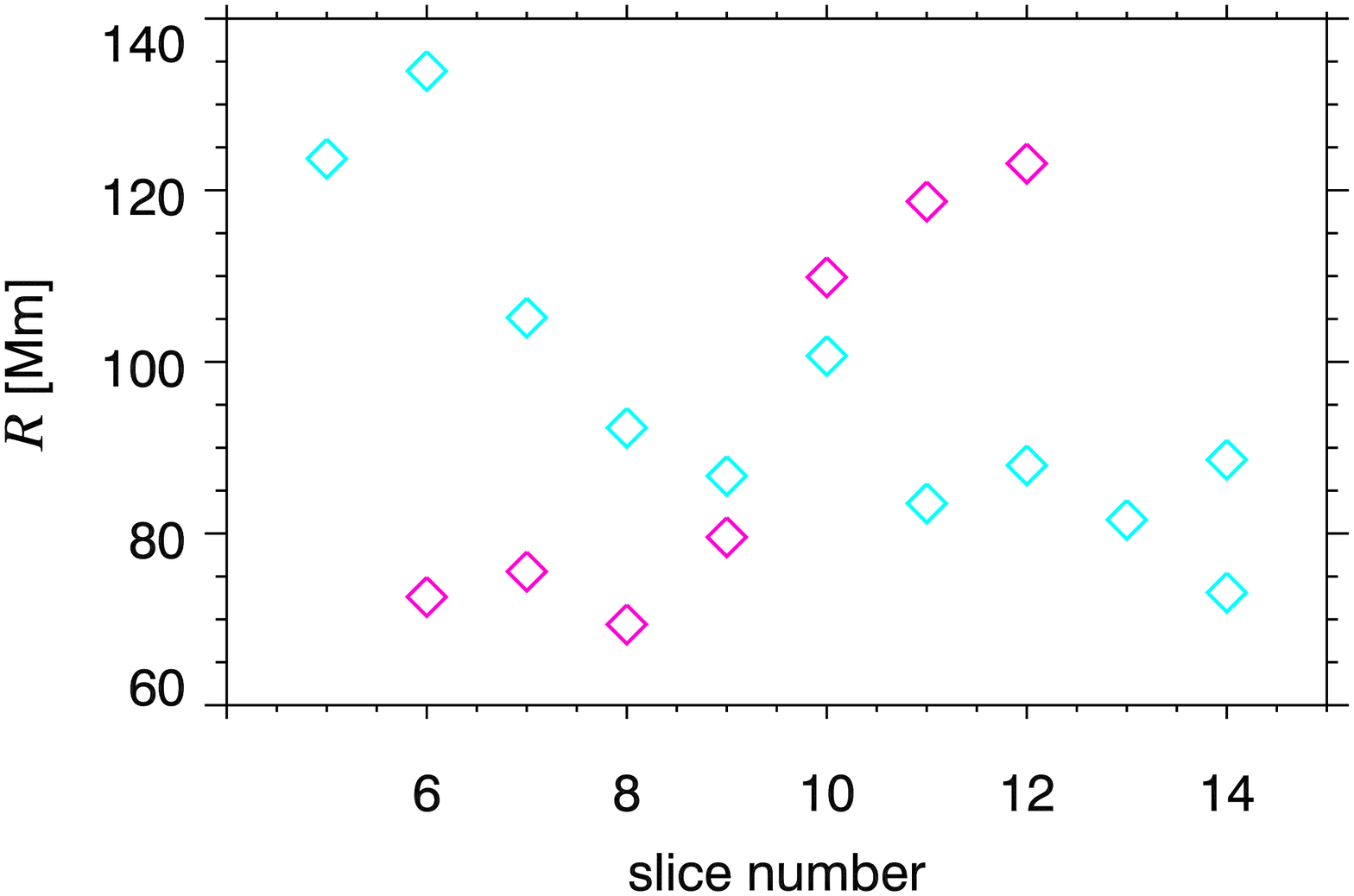}
\caption{Estimated curvature radii ($R$) of the magnetic dips supporting the filament according to the observed periods.
The cyan and magenta symbols represent the results before and after the mass drainage.
\label{fig15}}
\end{figure}

As to the restoring forces of the longitudinal filament oscillations, there are a few explanations, such as the magnetic pressure gradient along the filament axis \citep{vrs07}, magnetic tension and gravity \citep{li12}, magnetic 
pressure and gravity \citep{shen14b}. Both analytical solutions and numerical simulations have proven that the dominant restoring force is the gravity of filament along the magnetic dip 
\citep{luna12a,luna12c,zqm12,zqm13,ter15}. According to the formula ($P=2\pi \sqrt{R/g_{\odot}}$) and the periods of oscillations in Table~\ref{tab:fit}, the curvature radii of the magnetic dips supporting the filament 
can be estimated. Figure~\ref{fig15} shows the results of estimated values (69.4$-$133.9 Mm), which are larger than the values (43$-$66 Mm) reported by \citet{luna14}.
Besides, according to the formula ($B_{tr} [G]\geq17P [hr]$) in \citet{luna14} and the periods in Table~\ref{tab:fit}, 
the lower limit of the transverse magnetic field strength of the dips is estimated to be 15 G, which is close to the values reported by \citet{luna14} and \citet{bi14}.

\subsection{Why do the amplitudes of oscillations damp and grow?}
Compared to the restoring force, the damping mechanism of longitudinal oscillations is poorly understood. Using the 1D HD numerical simulations, \citet{zqm13} found that the major damping mechanism is optically thin
radiative loss, although the role of thermal conduction could not be ignored. In the case of extremely large amplitudes so that part of the filament material drain into the chromosphere along one of the leg, the remaining part 
continues to oscillate, whose damping timescale is significantly shortened. On the other hand, gradual mass loading is believed to play an important role in the fast damping of longitudinal oscillations \citep{luna12a,luna14,rud16}.
It is not easy to distinguish the two mechanisms from observations, such as $\tau/P$. However, considering that the filament material is supported by the magnetic tension force of the dip, continuous mass accretion may reduce 
the curvature radius of the dip and the period of oscillation accordingly. In our study, the periods nearly keep constant both before and after the mass drainage (see Figure~\ref{fig12}). Hence, we incline to the believe that the 
longitudinal oscillations are damped via radiative loss of the coronal and transition region plasmas surrounding the filament.

\begin{figure*}
\epsscale{1.0}
\plotone{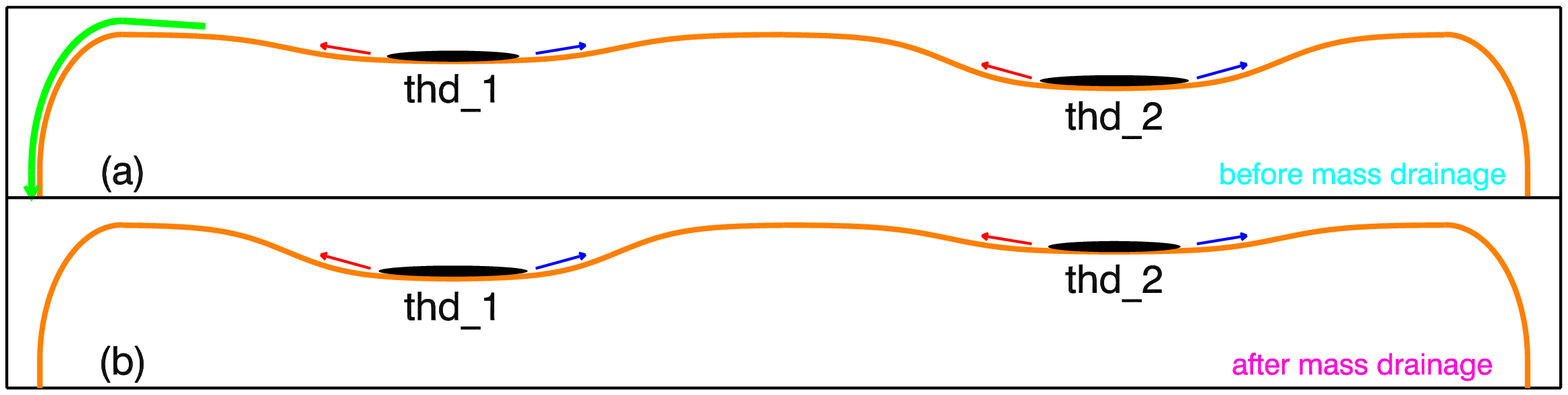}
\caption{A schematic cartoon to illustrate the thread-thread interaction during the filament oscillations before and after the mass drainage. 
The two filament threads (thd\_1 and thd\_2) are supported by two magnetic dips that have different curvature radii.
The green arrow represents the direction of mass drainage to the photosphere (see text for details).
\label{fig16}}
\end{figure*}

Surprisingly, we observed amplitude growth with time, i.e., negative $\tau$, which has never been reported and explored. 
The growth of amplitude in coronal loop oscillations is equally rare. For the first time, \citet{wang12} observed growing transverse oscillations of a multistranded coronal loop with period of $\sim$230 s. 
The growing times for the upper and lower strands are 1248 s and 759 s. The growing ratio ($\tau/P$) are 5.43 and 3.26, respectively. The authors proposed that the amplitude-growing kink oscillations may result from continuous 
non-periodic driving by magnetic deformation of the CME. In our study, the growth occurred at certain positions and certain times (see Figure~\ref{fig12} and Figure~\ref{fig14}(c)), and most of the growing ratios lie in the range of
-1.5 to -9. Following the previous work \citep{zqm13}, \citet{zhou17} conducted 1D radiative HD numerical simulations of longitudinal filament oscillations when two threads are magnetically connected. The thread-thread
interaction is investigated in detail. In case A (see their Fig. 3), the displacement of the active thread can be perfectly described by an exponentially decreasing sine function. The amplitude of the passive thread, however, 
increases rapidly during the first 85 minutes before decreasing gradually. Since the initial perturbation is imposed in the active thread, the energy is transferred from the active thread to the passive thread by sound waves (see 
their Fig. 2), which leads to the rapid growth of amplitude of the passive thread. As time goes on, the transferred energy is overtaken by the radiative loss and thermal conduction. Therefore, the amplitude of the passive thread
begins to decrease. In our study, the oscillations along S5, S6, S7, and S8 are somewhat similar to the situation of the passive thread (see Figure~\ref{fig13}), implying that thread-thread interaction may exist during the oscillations.
To explain the complex behaviors of the longitudinal filament oscillations, we tentatively propose a schematic cartoon by introducing the thread-thread interaction. In Figure~\ref{fig16}, the two filament threads (thd\_1 and thd\_2)
are supported by two magnetic dips with different curvature radii. The oscillations of thd\_1 represent the oscillations along S5-S8, and the oscillations of thd\_2 represent the oscillations along S9-S14.
When the filament is disturbed by a subflare or microflare in the filament channel, it starts to oscillate. Before mass drainage, the periods and curvature radii of thd\_1 are larger than those of thd\_2. 
The oscillation of thd\_2 is active, while the oscillation of thd\_1 is passive or driven, since the amplitudes of S9$-$S14 damp with time, while the amplitudes of S5-S8 grow with time. Besides, 
the starting times of oscillations of S5-S8 seem to be delayed compared with S9-S14, which is consistent with the result of simulations. The mass drainage plays a key role in altering the magnetic 
configuration of the filament, so that the roles of threads at different places change. It is likely that the southward mass drainage results in decrease (increase) of curvature radii of thd\_1 (thd\_2), respectively.  After mass drainage, 
the situations change significantly. The periods and curvature radii of thd\_1 are lower than 
those of thd\_2. The oscillation of thd\_1 is active, while the oscillation of thd\_2 is passive, since the amplitudes of S5-S8 damp with time, while the amplitudes of S9-S14 grow with time. 
In the next step, we will investigate the thread-thread interaction during the longitudinal filament oscillations with mass exchange included.

\section{Summary} \label{sec:summary}
In this paper, we report our multiwavelength observations of the large-amplitude longitudinal filament oscillations on 2015 May 3 using various instruments. The main results are summarized as follows.
\begin{enumerate}
\item{The sigmoidal filament was located next to AR 12335. Large-scale magnetic configuration and the EUV images in 171 {\AA} reveal that it was constrained by the overlying arcade.}
\item{The filament oscillations were most probably triggered by the magnetic reconnection in the filament channel, which is characterized by the bidirectional flows, brightenings in EUV and SXR, and magnetic cancellation 
in the photosphere.}
\item{The directions of oscillations have angles of 4$^\circ$-36$^\circ$ with respect to the filament axis. The whole filament did not oscillate in phase as a rigid body. Meanwhile, the periods of oscillations, ranging from 3100 s to 
4400 s, have a spatial dependence, implying that the curvature radii of the magnetic dips are different at different positions. The values of $R$ are estimated to be 69.4$-$133.9 Mm, and the minimum transverse magnetic field of 
the dips is estimated to be 15 G. The amplitudes of S5-S8 grew with time, while the amplitudes of S9-S14 damped with time. The amplitudes of oscillations range from a few to ten Mm, and the maximal velocity 
can reach 30 km s$^{-1}$.}
\item{Interestingly, the filament experienced mass drainage southwards at a speed of $\sim$27 km s$^{-1}$. The oscillations continued after the mass drainage and lasted for more than 11 hr. After mass drainage, the phases did 
not change a lot, which is consistent with the prediction of HD numerical simulation. The periods of S5-S8 decreased, while the periods of S9-S14 increased. The amplitudes of S5-S8 damped with time, 
while the amplitudes of S9-S14 grew. Most of the damping (growing) ratios are between -9 and 14.}
\item{We propose a schematic cartoon to illustrate the complex behaviors of oscillations by introducing thread-thread interaction. The positions of active and passive oscillations change after the mass drainage, 
which is possibly due to the magnetic configuration of the filament channel is altered.
Our results provide strong constraints to the future theoretical models of filament oscillations. Additional case studies and numerical simulations are urgently required to understand the nature of filament oscillations.}
\end{enumerate}

\acknowledgments
The authors appreciate the referee for valuable comments and suggestions. We also thank M. Luna, Y. H. Zhou, J. T. Su, D. Li, Y. Dai, and K. Yang for fruitful discussions. 
Q. M. Zhang acknowledges support from the International Space Science Institute (ISSI) to the Team 314 on ``Large-Amplitude Oscillation in prominences" led by M. Luna.
\textit{SDO} is a mission of NASA\rq{}s Living With a Star Program. AIA and HMI data are courtesy of the NASA/\textit{SDO} science teams. 
This work is supported by the Youth Innovation Promotion Association CAS, NSFC (Nos. 11333009, 11533008, 11603013, 11473071, 11573072), the Fund of Jiangsu Province (Nos. BK20161618, BK20161095, and BK20141043), 
CAS Key Laboratory of Solar Activity, National Astronomical Observatories (KLSA201716), and the One Hundred Talent Program of Chinese Academy of Sciences.

\end{document}